# Suppression of Phase Separation in AlGaInAs Compositionally Graded Buffers for 1550 nm Photovoltaic Converters on GaAs

Kevin L. Schulte,[a)] John F. Geisz, Harvey L. Guthrey, Ryan M. France, Edgard Winter da Costa, and Myles A. Steiner

*National Renewable Energy Laboratory, Golden, Colorado 80401, USA*

[a)]**Author to whom correspondence should be addressed**: kevin.schulte@nrel.gov

**ABSTRACT**

We investigate strategies to suppress phase separation and reduce threading dislocation densities (TDD) in AlGaInAs compositionally graded buffers (CGBs) that span the lattice constant range from GaAs to InP. Combining results from high resolution x-ray diffraction, cathodoluminescence, transmission electron microscopy, and photovoltaic device measurements, we correlate choices of epitaxial growth conditions with the defect structure of the CGBs and subsequent device performance. Both the use of substrates with high misorientation off (100) towards the (111)A plane and Zn-doping instead of Si-doping are shown to suppress phase separation and reduce TDD. We demonstrate a 0.74 eV GaInAs device grown on a (411)A GaAs substrate using a Zn-doped AlGaInAs CGB with TDD = $3.5 \pm 0.2 \times 10^6$ cm$^{-2}$ that has a bandgap-open circuit voltage offset of only 0.434 V measured under the AM1.5G solar spectrum. We characterized this device under high-intensity irradiance from a 1570 nm laser and measured a 31.9% peak efficiency laser power conversion efficiency at 3.6 W/cm$^2$. These results provide a roadmap to the manufacture of laser- and thermal-power conversion devices with the performance and cost-effectiveness needed to drive adoption of these technologies at scale.

## I. INTRODUCTION

Power beaming applications require photovoltaic laser power converters (LPCs) to convert photonic energy sources (e.g. lasers) to electric current with high efficiency. 1550 nm is a particularly interesting wavelength for this application due to the availability of sources and converters, the existence of an atmospheric window for wireless applications, and low attenuation by optical fibers for wired applications.[1] Devices near this wavelength are also of interest to thermophotovoltaic applications that convert radiation from emitters with temperatures around 1500 °C.[2,3] 0.74 eV GaInAs grown lattice-matched to InP offers the highest performance to date, however, growth on other substrates such as GaAs or Ge offers the potential for significant cost reduction because these substrates cost less and are available in larger diameters, enhancing scalability. Growth on these substrates requires the use of strategies such as compositionally graded buffers (CGBs), which bridge the lattice constant difference between the substrate and the GaInAs composition of interest while maintaining a low threading dislocation density (TDD).[4]

Photovoltaic applications ideally target a TDD of $1 \times 10^6$ cm$^{-2}$ or below.[5] CGBs using Ga$_z$In$_{1-z}$P provide one solution to bridge between GaAs and InP with demonstrated TDD near the target and high performance in photovoltaic applications.[6,7] Al$_y$Ga$_{1-x-y}$In$_x$As CGBs[8-11] present another option that is potentially more cost effective, because they utilize less In, which is historically more expensive than Ga or Al, and do not require as high of group V overpressure that $Ga_zIn_{1-z}P$ CGBs do.[12] $Al_yGa_{1-x-y}In_xAs$ on GaAs CGBs are grown with similarly low TDD to $Ga_zIn_{1-z}P$ in the range of $0 < x < 0.3$, which access GaInAs bandgaps of 1.0-1.4 eV, but usually contain elevated TDD above this In content.[6,8,9,11,13-15] The increase in TDD is attributed to occurrence of phase separation above this composition, which creates stress fluctuations in the CGB that inhibit dislocation glide and annihilation. The use of substrates with misorientation towards (111)A helps suppress phase separation, leading to reduced TDD.[11,15] Sun *et al.*[11] demonstrated a TDD of 7 x 10$^6$ cm$^{-2}$ in an AlGaInAs CGBs from GaAs to InP grown on GaAs with a 15° misorientation towards (111)A.

Previous work also suggests that dopant choice impacts the defect structure of III-V CGBs. Tangring et al.[16] showed that p-type Be doping led to improved GaInAs CGB roughness and TDD compared to n-doped and undoped structures. Fan et al.[17] found that p-type GaAsP CGBs exhibited lower TDD compared to undoped and n-type CGBs. These results are consistent with studies demonstrating that dislocation glide velocities in III-V semiconductors are strongly dependent on the dopant choice and type.[18,19] In particular, Zn promotes high β-dislocation velocities in GaAs, especially relative to n-type dopants like Te.[19] Dislocation velocity is inversely correlated with TDD in theory,[20] meaning that higher dislocation velocities should reduce TDD. He et al.[21] showed that Zn doping improved the roughness and x-ray diffraction peak width in InP layers grown on AlGaInAs CGBs grown on 15°A GaAs substrates, but the TDD was not reported.

In this work, we explore the effect of substrate offcut and dopant type on the microstructure of AlGaInAs CGBs from GaAs to InP. We find that a high misorientation off (100) towards (111)A and Zn doping are effective at reducing phase separation and TDD and maximizing photovoltaic device performance. We develop inverted growth of ~1550 nm GaInAs laser power converters using a removable Zn-doped grade, enabling the use of an n-on-p device configuration. We characterize these devices using a high-intensity 1570 nm laser diode source and find that use of optimized AlGaInAs growth conditions (411A substrate, Zn-doped grade) yields peak LPC efficiency of 31.9% at 3.6 W/cm$^2$, a 59% relative increase in peak efficiency compared to a similar device grown on a (100) 6°A substrate with a Si-doped grade.

## II. EXPERIMENTAL METHODS

All materials were grown by atmospheric pressure organometallic vapor phase epitaxy (OMVPE). Devices were grown on Si-doped (100) GaAs substrates with a misorentation of 6° or 19.5° towards the (111)A plane. The latter substrate corresponds to a (411)A plane. Two types of devices were studied. Fig. 1(a) shows a GaInAs/AlInAs double heterostructure (DH) used for structural characterization. This device is grown in an upright configuration using a step-graded $Al_{0.50}Ga_{1-x}In_xAs$ CGB spanning from GaAs to $Al_{0.43}In_{0.57}As$. The compositional steps are 0.25 μm thick, and use a misfit grading rate of 0.8%/μm. We note that the Al content was held at 0.5, with the Ga content decreased as the In content was increased until the Ga content reached 0% at which point the Al was decreased in order to continue adding In. A 0.5 μm strain-overshoot layer was the final grading layer, and then the DH was grown lattice matched to the in-plane lattice constant of the overshoot layer with residual strain. The stress in the growing film was monitored with an *in situ* laser curvature measurement. The CGB growth temperature was 720 °C with a growth rate of 5 μm/h and a V/III ratio of 300, conditions intended to suppress phase separation.[13] The GaInAs layer and the AlInAs cladding layers were grown at 620 °C. Four DHs were grown, one with CGB doping of 1 x $10^{18}$ cm$^{-3}$ Si and one with 1 x $10^{18}$ cm$^{-3}$ Zn on each of the 6°A and (411)A substrates. The DH doping was the same type as in the CGB in each case. Figs. 1(b) and (c) show the inverted GaInAs photovoltaic cell structures used in this study. These devices were grown with the same CGB structure as the DHs, followed by a GaInAs n-on-p cell passivated with InP on the front and back. Cells were grown on Si-doped, Zn-doped, and undoped CGBs, with nominal emitter doping of 1 x $10^{18}$ cm$^{-3}$ Se and nominal base doping of 4 x $10^{16}$ cm$^{-3}$ Zn. The Si-doped CGB devices have a baseline structure with the GaInNAs:Se front contact coming before the CGB, with the CGB left in the final device, as in Fig. 1(b). The structures with Zn-doped or nominally undoped grades required removal of the CGB to avoid forming a second p-n junction in opposition to the photovoltaic cell, motivating use of the structure of Fig. 1(c). This structure contains a GaInAs:Se front contact layer grown just before the cell,[22] with an InP etch stop inserted just after the grade to enable removal of the grade and place the front contact directly on the n-side of the cell. Devices were processed into 0.116 cm$^2$ square mesas with a concentrator solar cell grid as described in ref. 23. After processing, a bilayer ZnS/MgF$_2$ antireflection coating was deposited on select photovoltaic devices by thermal evaporation.

Multiple structural characterization techniques were applied to the DHs. Atomic force microscopy (AFM) in tapping mode was used to measure the surface roughness. High resolution x-ray diffraction was used to measure (400) reciprocal space maps of the DHs in the [011] and [0-11] azimuthal directions. Dislocation density was analyzed using cathodoluminescence generated using a scanning electron microscope as described in ref. 23 An area of at least 2 x $10^{-4}$ cm$^2$ was analyzed for each sample. The DH samples were also analyzed using scanning transmission electron microscopy (STEM). (1-10) cross-section specimens were prepared using a focused ion beam. Images were collected using a high angular annular dark field detector, which yields images in which variations in contrast are proportional to variations in the local atomic weight of the sample.

External quantum efficiency and light current density-voltage (*J-V*) measurements were performed on the devices as described in ref. 24. The Xe lamp intensity to simulate the AM1.5 Global (AM1.5G) spectrum was set using a calibrated 0.74 eV GaInAs cell with an InP filter. *J-V* curves were also measured under monochromatic 1570nm laser illumination over a wide range of irradiances with the stage temperature controlled to 25°C. The irradiance of the laser was varied by applying different driving currents to the laser as well as using neutral density filters in front of the laser just above the threshold driving current. The average irradiance ($E_{tot}$) of the device was calculated from the short-circuit current ($J_{SC}$) as described in Geisz et al.[25] The illuminated area of 0.100 cm$^2$, excluding the busbar area, was used for LPC measurements.

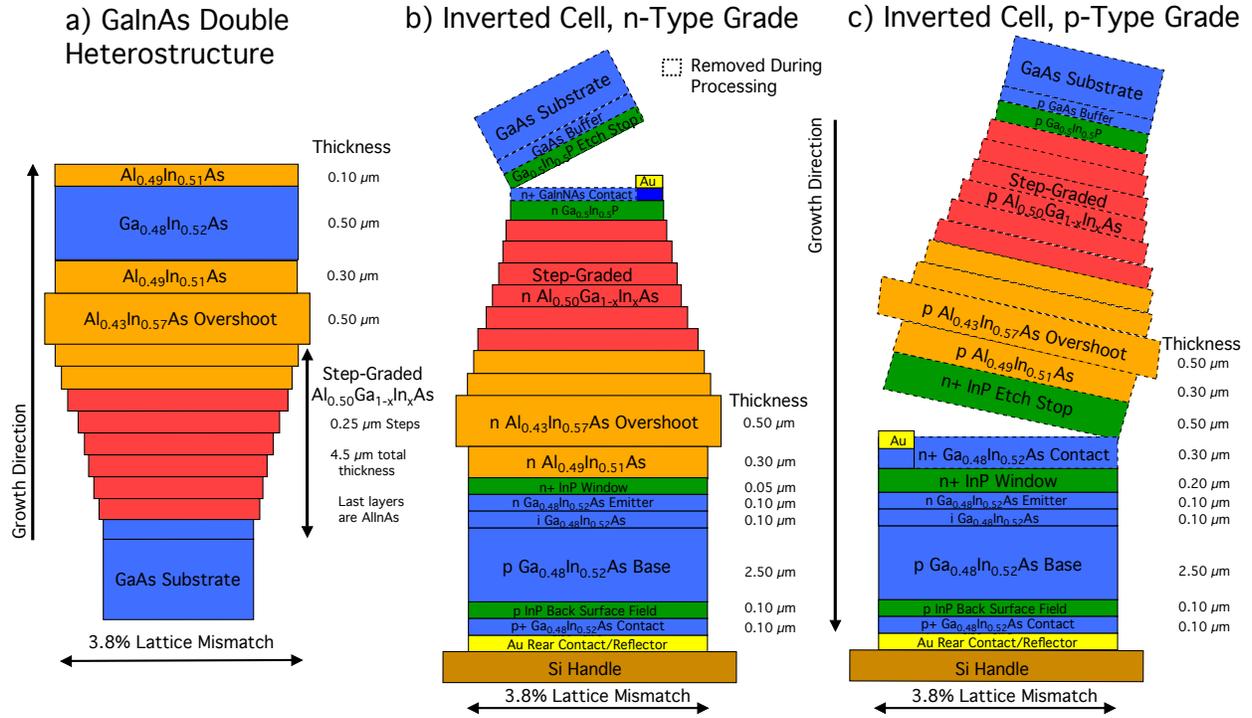

Fig. 1.  Device structures used in this study. a) GaInAs/AlInAs double heterostructure grown upright on GaAs using an AlGaInAs CGB. b) Inverted n-on-p GaInAs photovoltaic cell grown on GaAs using an n-type AlGaInAs CGB. The device is shown after processing and bonding to an Si handle. The CGB remains within the device. c) Inverted n-on-p GaInAs photovoltaic cell grown using a Zn-doped or nominally undoped CGB. The grade is removed from the structure during processing.

## III. RESULTS

First, we characterized the surfaces of the four DH samples using AFM. Fig. 2 shows 50 μm x 50 μm height maps of each sample surface, with the root mean square (RMS) roughness listed in each panel. Looking at the 6°A samples, the surface of the Si-doped grade exhibits the typical cross-hatch roughness observed in III-V compositional grades, with an RMS roughness of 48.4 nm. The use of Zn-doping in the grade leads to a dramatic reduction in the roughness, down to 22.4 nm. The (411)A samples also exhibit cross hatch, but with larger asymmetry between the cross-hatch lines due to the higher inclination of the active dislocation slip systems relative to the surface normal. Both doping types yield similar roughness values, which are in between those of the two 6°A samples.

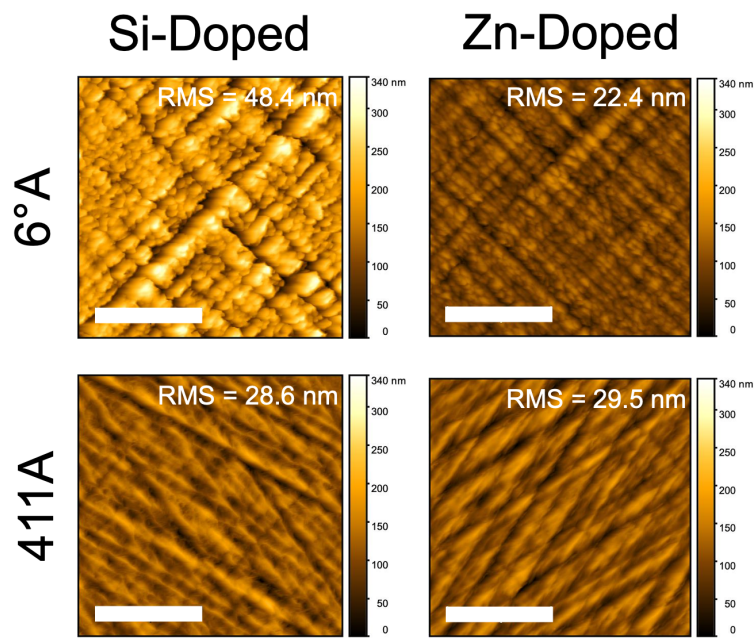

Fig. 2.   50 µm x 50 µm AFM scans of the surface of the double heterostructures. The root mean square roughness is listed in each image. Scale bar indicates 20 µm.

We used high resolution reciprocal space mapping to analyze the crystal structure of the DH samples. Fig. 3 shows (400) reciprocal space maps of each sample in the orthogonal [011] and [0-11] azimuthal directions. The intensity is arbitrary and plotted using a logarithmic scale. The full width at half maximum (FWHM) and tilt in the omega direction of the GaInAs reciprocal lattice points in Fig. 3 are listed in Table 1. The FWHM is correlated in theory with defects in the crystal such as threading dislocations or phase separation.[26,27] The tilt provides some information about the strain relaxation mechanism via dislocations. Strain relaxation predominantly occurs in III-V materials by formation of 60° dislocations on 111 glide planes. These dislocations contain a burgers vector component that causes tilt, and the sense of the tilt can be positive or negative. The tilt is low when the population of dislocations is balanced with positive and negative tilt vectors, but the use of an offcut is known to cause negative tilting in the offcut direction, by selecting a specific dislocation type due to an imbalance in the shear stress on the available glide planes.[28] This phenomenon is the reason all of the samples show a large negative tilt in the [011] or "A" direction; furthermore, the magnitude of the tilt is proportional to the dislocation selectivity.[7] The 6°A Si-doped sample has the highest FWHM in both azimuthal directions, possibly indicating a high degree of defects and/or phase separation. Use of Zn-doping on the 6°A sample leads to a pronounced reduction in FWHM, particularly in the [011] scan direction. The reciprocal lattice points of the AlGaInAs compositional steps up to the final layer also become narrower, whereas in the Si-doped sample those points broaden significantly at about -3000 arcsec Omega-2Theta, which could indicate the occurrence of phase separation. The Si-doped (411)A sample has a significantly lower FWHM than the 6°A Si-doped sample, and lower [011] tilt. The use of Zn doping does not seem to affect the FWHM or tilt magnitude on the (411)A substrate.

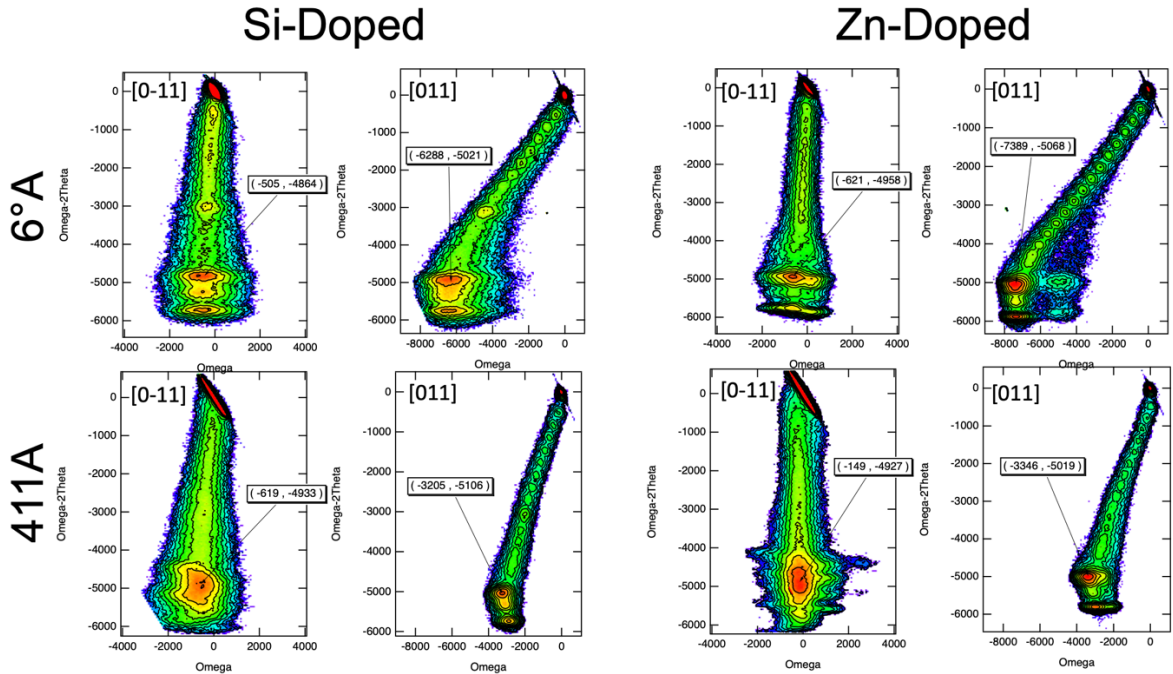

Fig. 3.  (400) reciprocal space maps of the GaInAs double heterostructures taken in the orthogonal [011] and [0-11] azimuthal directions. The Omega and Omega-2Theta coordinates of the reciprocal lattice point of the GaInAs layer are listed in each figure.

Table 1. Full width half maximum and tilt in the omega direction of the GaInAs reciprocal lattice point of the RSMs in Fig. 3. Units in arcseconds.

|  | 6degA | | | | 411A | | | |
|---|---|---|---|---|---|---|---|---|
|  | [0-11] | | [011] | | [0-11] | | [011] | |
|  | FWHM | Tilt | FWHM | Tilt | FWHM | Tilt | FWHM | Tilt |
| **Si-Doped CGB** | 1634 | -505 | 1934 | -6288 | 882 | -619 | 721 | -3205 |
| **Zn-Doped CGB** | 1429 | -515 | 840 | -7389 | 847 | -149 | 791 | -3346 |

We next employed CL imaging to probe for dislocations in the DHs. Fig. 4 shows the secondary electron images and panchromatic CL images for the four structures. The threading dislocations appear as black spots due to reduced luminescence. The TDD calculated from each sample is listed below each image. The Si-doped 6°A sample had a large concentration of dark pixels with so much overlap that it was not possible to calculate the TDD. The Zn-doped 6°A sample had a lower density of dark pixels and discrete dark spots. The TDD of $5.9\pm0.2 \times 10^6$ cm$^{-2}$ is clearly lower than the Si-doped case. Use of the (411)A substrate yielded a TDD of $6.4\pm0.3 \times 10^6$ cm$^{-2}$ in the Si-doped sample, which decreased to $3.5\pm0.2 \times 10^6$ cm$^{-2}$ in the Zn-doped case. We also used TEM to analyze the defect structure of these samples. Fig. 5 shows high angle annular dark field images of the (0-11) cross sections of each sample taken in a STEM. The Si-doped 6°A sample is shown in panel (a). We note that the interfaces between the layers exhibit significant undulation. Also, there is contrast indicating some long-period (100s of nm) phase separation in the compositional steps and in the upper AlInAs layers. An instance in which a dislocation appears to be pinned by a region of phase separation is indicated by the "A". Dislocation pinning is one of the primary mechanisms by which phase separation increases TDD.[6]

There is also shorter-period (10s of nm), roughly [100]-aligned phase separation in the first AlInAs cladding layer grown after the overshoot layer (see Fig. 1) indicated by the "B". The AlInAs layers in the DH were grown at a lower temperature than the compositional steps in the CGB (620 °C vs. 720 °C), a parameter which has previously been correlated with phase separation in AlInP.[29] A growth pause was not used for this transition, meaning that the initial portion of the layer (which is free of phase separation) is grown hotter than 620 °C. We note that the sample cooled 50% of the way down to the new setpoint after 0.50 min and 100% of the way down after 1.15 minutes. These data suggest that growth temperature is one key factor to controlling phase separation in AlInAs. The vertically-aligned phase separation is not present in the Zn-doped 6°A sample in panel (b) of Fig. 5, suggesting that the Zn-doping has suppressed it, although some unidentified contrast remains in that layer and the interfaces exhibit significant undulation. There is still evidence of phase separation in the compositional steps, meaning that the Zn-doping has not suppressed all the phase separation. In the Si-doped (411)A sample [panel (c)], the short period phase separation is present in the cooler AlInAs layer, but it is fainter than in the 6°A sample, and inclined away from [100]. This inclination is presumably related to the significant misorientation of the substrate off (100). There appears to be some phase separation in the GaInAs layer as well, appearing as a dark stripe. The interfaces are significantly flatter than in either of the 6°A samples. There is no short period phase separation observed in the AlInAs layers of the Zn-doped (411)A sample [panel (d)], although there may still be a small indication of phase separation in the GaInAs layer. The interfaces between the layers exhibit good flatness.

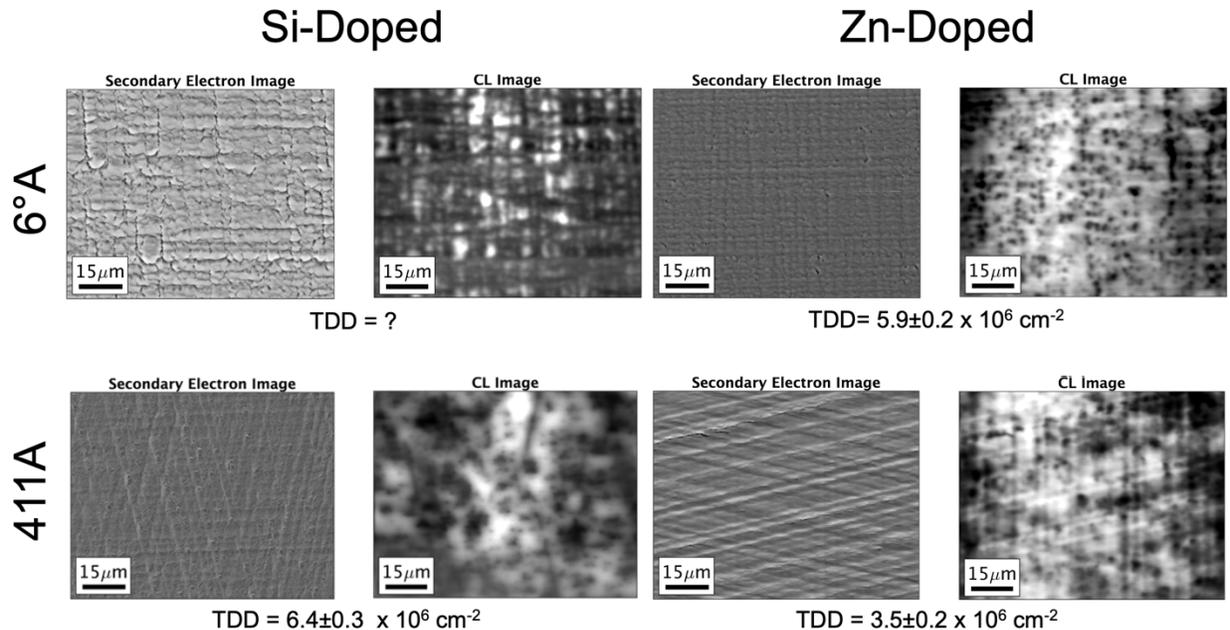

Fig. 4. Scanning electron microscopy and panchromatic cathodoluminescence images of the double heterostructures. The threading dislocation density is listed below each image.

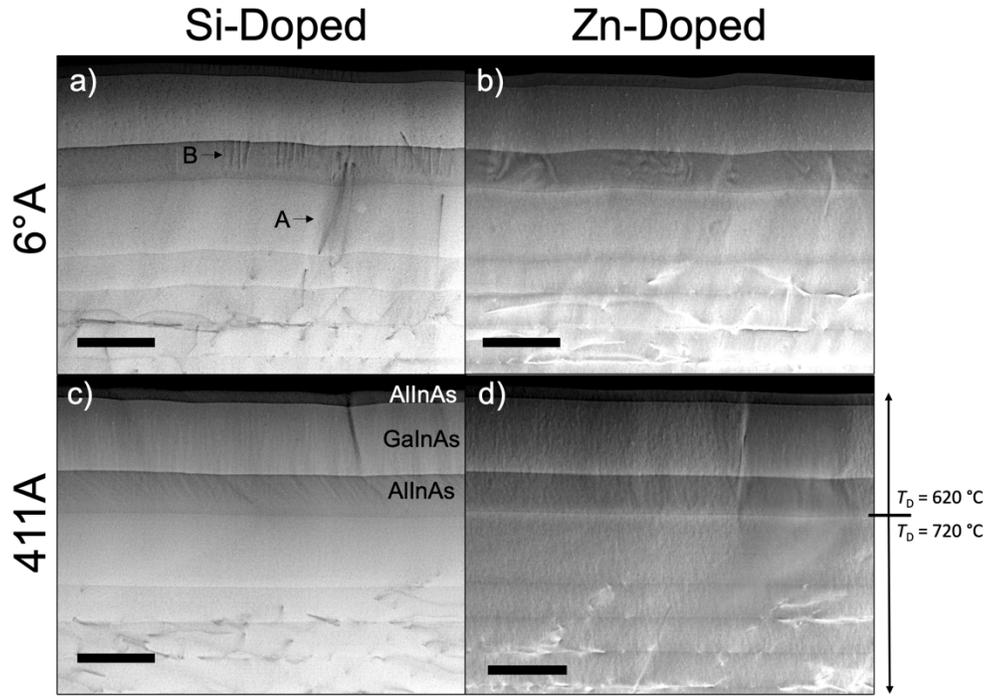

Fig. 5. (0-11) cross-section STEM high angular annular dark field images of the four double heterostructure samples. The scale bar indicates 500 nm. The double heterostructure layers are labeled in c), and a temperature change that occurred during the growth for all samples is indicated by the arrows in d).

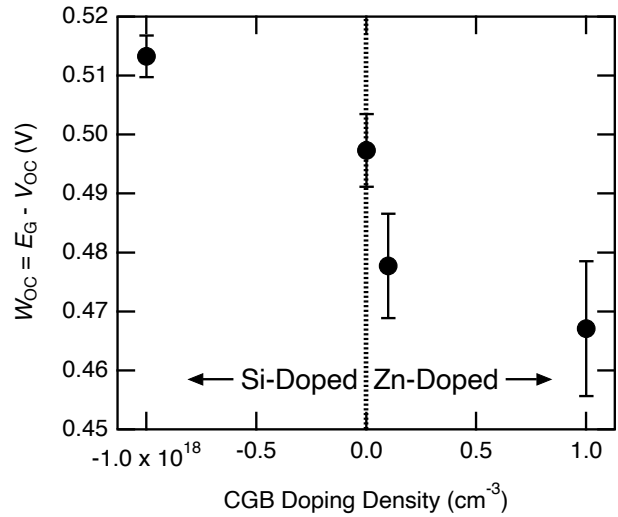

Fig. 6. Bandgap-voltage offset ($W_{OC} = E_G/q - V_{OC}$) of 0.74 eV GaInAs grown on 411A substrates with varying dopant density in the compositionally graded buffer layer. Negative values indicate n-type (Si-doped) and positive values indicate p-type (Zn-doped). Error bars indicate standard deviation over 8-11 devices measured for each sample.

We then grew 0.74 eV GaInAs cells on AlGaInAs CGBs on (411)A substrates with varying doping type and density in the CGB to test the effect of this parameter on the CGB quality. These devices were measured under the AM1.5G spectrum and we calculated the bandgap-voltage offset ($W_{OC}$) from the equation $W_{OC} = E_G/q - V_{OC}$, where $E_G$ is the device bandgap, q is the elementary charge, and $V_{OC}$ is the open circuit voltage. This metric is an indicator of material quality due to the fact that $V_{OC}$ is a function of recombination in the device and strongly correlated with TDD.[5] Fig. 6 plots $W_{OC}$ as a function of doping density. These devices did not have anti-reflection coatings. Note that negative values indicate n-type doping with Si, while positive values indicate p-type doping with Zn, and a value of zero indicates a nominally undoped grade. $W_{OC}$ is 0.513±0.004 V in the device with n = 1x10$^{18}$ cm$^{-3}$ (Si) doping, then decreases by 10 mV in a sample grown with an undoped CGB (doping density = 0). The $W_{OC}$ decreases further when the dopant is Zn, reaching a value of $W_{OC}$ = 0.478±0.009 V at p = 1x10$^{17}$ cm$^{-3}$ then decreases further with Zn doping down to $W_{OC}$ = 0.467 ±0.011 V for p = 1x10$^{18}$ cm$^{-3}$.

Next, we compared a 0.74 eV GaInAs laser power converter grown on the most optimal combination of (411)A GaAs substrate with 1x10$^{18}$ cm$^{-3}$ Zn-doped grade [see Fig. 1(b)], with a device grown on the least optimal combination of (100) 6°A substrate with an 1x10$^{18}$ cm$^{-3}$ Si-doped grade. Fig. 7 shows external quantum efficiency and light J-V measurements taken under a AM1.5G spectrum. Note that these devices have a bi-layer antireflection coating on the front surface. The baseline device has lower EQE at short wavelengths because it is filtered by the grade, whereas the device grown with the (411)A substrate had its buffer removed. The EQE of the baseline device peaks at 0.80, and then decreases monotonically with increasing wavelength. The (411)A device peaks at an EQE of 0.87, and then decreases more gradually with wavelength compared to the baseline. This trend suggests that the minority carrier diffusion length is larger in the (411)A device due to the significant reduction in TDD.[5] Looking at the *J-V* curves [Fig. 7(b)], the $V_{OC}$s of the baseline and (411)A devices are 0.217 and 0.312 V respectively, corresponding to $W_{OC}$ = 0.528 and 0.434 V, respectively. $W_{OC}$ = 0.4 ≤ is generally regarded as a desirable threshold for high-efficiency solar cells.[30] The optimized device compares favorably to 0.74 eV GaInAs metamorphic devices grown on GaInP.[31]

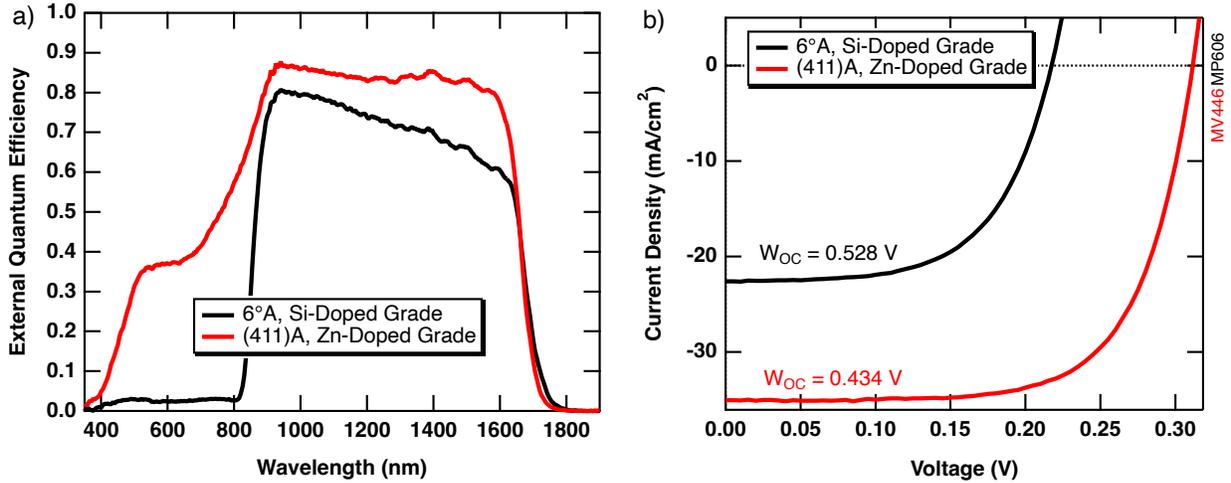

Fig. 7. (a) External quantum efficiency, and (b) light current density-voltage curves under AM1.5G measured from 0.74 eV GaInAs photovoltaic devices grown on a GaAs (100) 6°A substrate with an Si-doped graded buffer and a GaAs (411)A substrate with a Zn-doped graded buffer.

Lastly, we characterized these devices using high-intensity emission from a 1570 nm diode laser to understand their performance as LPCs. We also compared these devices to a lattice-matched (LM) device grown on InP[3] and a lattice-mismatched (LMM) device grown on (100) GaAs using a GaInP compositionally graded buffer with an estimated TDD of 2x10$^6$ cm$^{-2}$.[31] The EQEs of these four devices are shown in Fig. 8(a), and the power dependent $V_{OC}$, fill factor, and LPC efficiency are shown in Fig. 8(b). The optimized AlGaInAs CGB grown on (411)A GaAs shows a significant improvement in peak LPC efficiency relative to the (100) 6°A Si-doped CGB device, with a 59% relative increase from 21.2% at 8.0 W/cm$^2$ to 31.9% at 3.6 W/cm$^2$. We note that this device would benefit from some optimization to reduce series resistance, e.g. thickening of the InP window layer, which would help decrease the sheet resistance of the device and reduce the fill factor droop observed in this cell at high irradiance, as well as some optimization of the EQE at 1570 nm. The (100) 6°A GaAs, Si-doped CGB device still retains the CGB within the structure, which enables current spreading over the multiple μm of these layers, contributing to reduced FF droop and an efficiency that peaks at higher irradiance. The LMM cell grown with the GaInP CGB offers improved performance, peaking at 42.1% at an irradiance 7.0 W/cm$^2$, in part due to its lower TDD which enables increased $V_{OC}$ and improved current spreading due to the thick CGB. The LMM GaInP CGB device provides an attainable benchmark the AlGaInAs CGB devices if further improvements in TDD can be made. Lastly, the LM device grown on InP offers the best performance, as expected, with an efficiency of 47.4% at an irradiance of 3.2 W/cm$^2$.

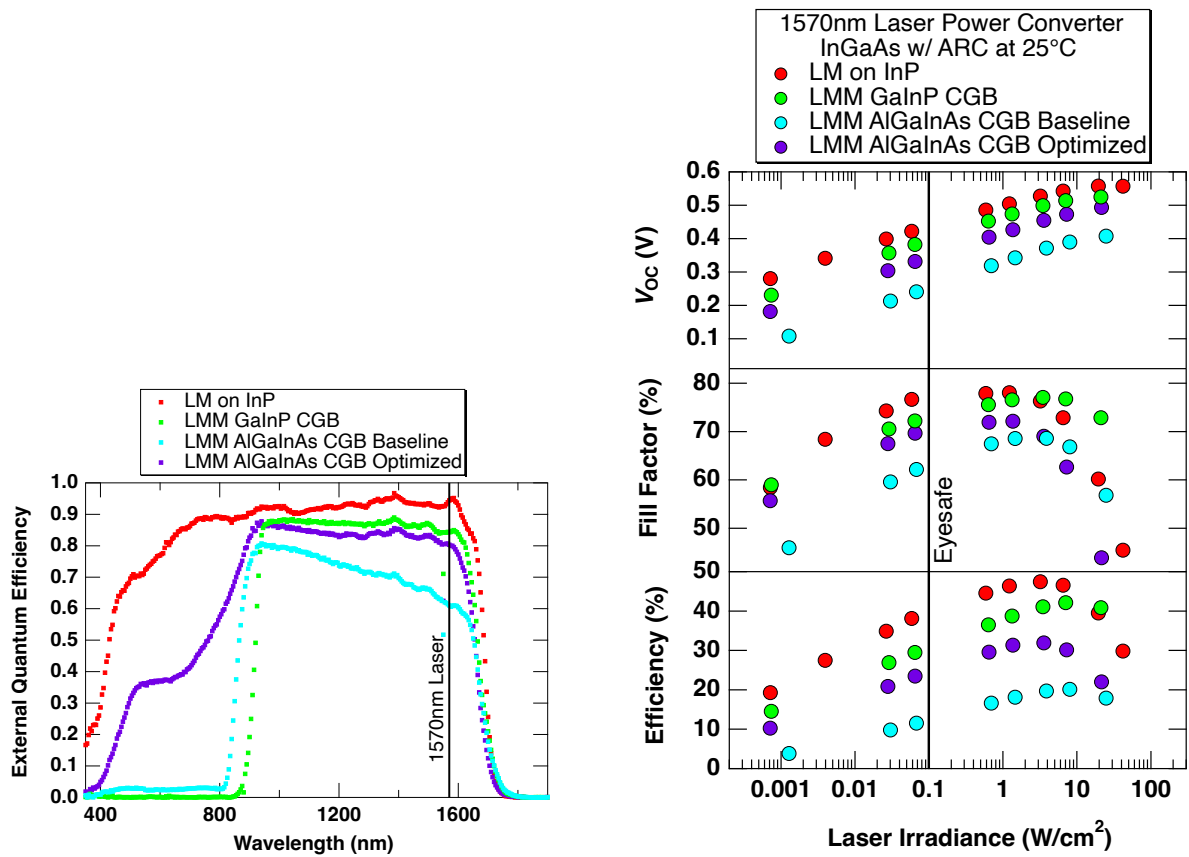

Fig. 8. 1570 nm laser power conversion measurements of four ~0.74 e GaInAs devices, (red) lattice-matched (LM) on InP, (green) lattice-mismatched on (100) GaAs using a GaInP CGB, (light blue) lattice-mismatched on (100) GaAs using an AlGaInAs CGB (baseline) and (purple) lattice-mismatched on (411)A GaAs using an AlGaInAs CGB

(optimized). (a) External quantum efficiency. (b) $V_{OC}$, Fill Factor, and conversion efficiency as a function of 1570 nm laser irradiance. An irradiance of 0.1 W/cm$^2$ or lower is considered "eyesafe" at 1570nm.

## IV. DISCUSSION

Epitaxial III-V alloys exhibit phase separation, even in cases of nominally lattice-matched growth, when it becomes favorable to lower the energy at the growing surface through a combination of local lattice parameter fluctuation and surface undulation (roughening).[32-36] Metamorphic materials are particularly susceptible to phase separation, because mismatch stress is a key driver of phase separation,[36] and metamorphic materials invariably roughen in response to stress fields from intentionally introduced dislocations.[37] The ability of the material to form multiple phases requires surface diffusion of adatoms, because diffusion in the bulk is slow, and so successful methods to prevent phase separation (e.g., the use of low deposition temperatures[10]) hinder adatom diffusion. The use of an offcut substrate introduces a high density of steps which serve as adatom sinks and limit surface diffusion lengths. The use of a high misorientation off (100) towards (111)A was successful in this and previous[11,15] work because it aligns the direction of higher adatom diffusivity, [011],[38] perpendicular to the step edges, maximizing the effectiveness of this strategy. The (411)A substrate also led to significantly reduced undulation of the interfaces, which is correlated with lower phase separation.[36,39]

The mechanism by which Zn-doping of the CGB impacts phase separation and TDD is less clear. It is possible that Zn reduces the adatom diffusion length directly, although there is minimal evidence for this effect in the literature. On the other hand, there is a large body of evidence that Zn is correlated with fast self-diffusion in GaAs via the concomitant introduction of fast diffusing interstitial point defects.[40,41] Diffusion of Zn into the bulk was observed to cause intermixing of GaInAs/AlInAs multiple quantum wells into a uniform AlGaInAs layer via this mechanism.[42] Furthermore, studies of an analogously surface-driven phenomenon to phase separation, CuPt-type ordering of GaInP,[43] showed that Zn doping of GaInP can cause intermixing of a growing crystal via self-diffusion in the bulk. Kurtz et al.[44] showed that OMVPE-grown Zn-doped $Ga_{0.5}In_{0.5}P$ epilayers with hole concentrations of 1x10$^{18}$ cm$^{-3}$ exhibited a higher degree of CuPt order near the surface compared to positions deeper in the epilayer, whereas lightly doped samples exhibited a relatively constant, high degree of ordering throughout. These data imply that the ordering created at the surface is subsequently destroyed in the bulk during growth, presumably by a Zn-aided diffusion mechanism. Lee et al.[45] also showed that high Zn doping caused disordering of $Ga_{0.5}In_{0.5}P$ but did not significantly affect the surface roughness or step structure of their epilayers, providing further evidence that the ordering is destroyed sub-surface by bulk diffusion. It is plausible that a similar bulk diffusion mechanism eliminated or prevented the formation of the short-range phase separation indicated by the B in Fig. 5(a). It is unclear whether the phase separation never formed in the first place, or whether it formed and was subsequently destroyed by diffusion in the bulk. In the context of minimizing TDD it would be ideal if the phase separation never formed in the first place, because that would ensure the longest dislocation glide lengths. The lowest TDD observed in this study is still significantly higher than that observed in GaInP grades free of phase separation,[31] so perhaps there is still some impact of phase separation on dislocation glide in the AlGaInAs grades.

There are also potential benefits related to dislocation glide velocities enabled by Zn-doping that may be providing benefit apart from, or in addition to phase separation prevention/minimization. Previous studies[18,19] showed that β-dislocation (group III cores) velocities in GaAs doped with donors were multiple orders of magnitude smaller than in undoped GaAs, and that the same velocities were orders of magnitude higher in Zn-doped GaAs.[19] α-dislocation (group V cores) glide velocities for Si-doped and Zn-doped GaAs remained similar to those in undoped material.[18,19] Presumably then, the slow β-dislocations are limiting in the Si-doped grades whereas faster α-dislocations are limiting in the Zn-doped grades. The XRD FWHM data may provide evidence for this, as the FWHM in the [011] azimuthal scan direction (correlated with β-dislocations) of the 6°A Si-doped grade is high, and higher than the FWHM of the scan in the [0-11] azimuthal direction (correlated with α-dislocations). The FWHM in [011] is cut in half when Zn-doping is employed, to a value that is lower than that in the [0-11] direction, corresponding to the expected trend in glide velocities when switching from Si- to Zn-doping. We saw $W_{OC}$ improvements in the devices shown in Fig. 6 when switching from Si to undoped to Zn doped grades, a trend that agrees nicely with the trends in β-dislocation velocity trends observed by Yonenaga et al.[19] Dislocations with higher velocities can glide further, providing more strain relief per dislocation, and may be less prone to pinning by phase separation, both effects which would reduce TDD and in turn $W_{OC}$. Lee et al.[45] did not observe disordering in $Ga_{0.5}In_{0.5}P$ with Zn-doping to a hole concentration of $1 \times 10^{17}$ cm$^{-3}$, which may suggest that our $1 \times 10^{17}$ cm$^{-3}$ Zn-doped Zn doped grade (see Fig. 6) did not obtain benefit from phase separation prevention, but did benefit from improved dislocation glide velocities, and then the $1 \times 10^{18}$ cm$^{-3}$ doped grade benefited further from phase separation prevention. Further studies are needed to clarify these effects.

## V. CONCLUSION

We developed 0.74 GaInAs laser power conversion devices grown metamorphically on GaAs with AlGaInAs compositionally graded buffers. We determined that phase separation was correlated with high TDD in the graded buffers, and found that the use of a (411)A substrate and Zn-doping (instead of Si-doping), were effective to control phase separation and reduce TDD in the CGBs. Furthermore, 0.74 GaInAs photovoltaic devices fabricated on GaAs using a (411)A substrate and Zn-doped AlGaInAs CGB exhibited an over 100 mV increase in one-sun open circuit voltage, and 59% relative increase (from 21.2% to 31.9%) in 1570 nm LPC efficiency relative to devices grown on (100) 6ºA substrates with Si-doped grades. These optimized devices compare favorably to state of the art metamorphic 0.74 eV GaInAs devices grown using GaInP graded buffers, with significant room for optimization. These results offer another path towards scalable and lower cost LPC devices grown on GaAs.

**ACKNOWLEDGMENTS**

The authors would like to thank Michelle Young, Waldo Olavarria and Sarah Collins for device growth and processing, and Dale Batchelor at EAG Eurofins for preparation and imaging of the TEM samples. This work was authored by Alliance for Sustainable Energy, LLC, the manager and operator of the National Renewable Energy Laboratory for the U.S. Department of Energy (DOE) under Contract No. DE-AC36-08GO28308. This work was supported by the Laboratory Directed Research and Development (LDRD) Program at NREL. The views expressed in the article do not necessarily represent the views of the DOE or the U.S. Government. The U.S. Government retains and the publisher, by accepting the article for publication, acknowledges that the U.S. Government retains a nonexclusive, paid-up, irrevocable, worldwide license to publish or reproduce the published form of this work, or allow others to do so, for U.S. Government purposes.

**AUTHOR DECLARATIONS**

*Conflict of Interest Statement*

The authors have no conflicts to declare.

*Author Contributions*

Conceptualization, K.L.S.; Methodology, K.L.S and R.M.F.; Formal Analysis, K.L.S, J.F.G., and R.M.F.; Investigation, K.L.S, J.F.G, H.L.G., E.W.C..; Writing – Original Draft, K.L.S.; Writing – Review & Editing, All Authors; Funding Acquisition, M.A.S and K.L.S.

**DATA AVAILABILITY**

The data that support the findings of this study are available from the corresponding author upon reasonable request.